\documentclass[sigconf, nonacm]{acmart}
\usepackage{adjustbox}
\usepackage[symbol]{footmisc}
\usepackage{svg}
\usepackage{color}

\AtBeginDocument{%
  \providecommand\BibTeX{{%
    \normalfont B\kern-0.5em{\scshape i\kern-0.25em b}\kern-0.8em\TeX}}}

\begin{document}

\title{Transfer Ranking in Finance: Applications to Cross-Sectional Momentum with Data Scarcity}




\author{Daniel Poh}
\affiliation{%
  \institution{Department of Engineering Science, Oxford-Man Institute of Quantitative Finance, University of Oxford}
  \streetaddress{Eagle House, Walton Well Road, OX2 6ED}
  \city{Oxford}
  \country{United Kingdom}}
\email{dp@robots.ox.ac.uk}

\author{Stephen Roberts}
\affiliation{%
  \institution{Department of Engineering Science, Oxford-Man Institute of Quantitative Finance, University of Oxford}
  \streetaddress{Eagle House, Walton Well Road, OX2 6ED}
  \city{Oxford}
  \country{United Kingdom}}
\email{sjrob@robots.ox.ac.uk}

\author{Stefan Zohren}
\affiliation{%
  \institution{Department of Engineering Science, Oxford-Man Institute of Quantitative Finance, University of Oxford}
  \streetaddress{Eagle House, Walton Well Road, OX2 6ED}
  \city{Oxford}
  \country{United Kingdom}}
\email{stefan.zohren@eng.ox.ac.uk}

\renewcommand{\shortauthors}{Poh D., Zohren S. and Roberts S.}

\begin{abstract}
Modern cross-sectional strategies incorporating sophisticated neural architectures outperform traditional counterparts when applied to mature assets with long histories. However, deploying them on instruments with limited samples generally produces over-fitted models with degraded performance. In this paper, we introduce Fused Encoder Networks -- a hybrid parameter-sharing transfer ranking model which fuses information extracted using an encoder-attention module from a source dataset with a similar but separate module operating on a smaller target dataset of interest. This approach mitigates the issue of models with poor generalisability. Additionally, the self-attention mechanism enables interactions among instruments to be accounted for at the loss level during model training and inference time. We demonstrate the effectiveness of our approach by applying it to momentum strategies on the top ten cryptocurrencies by market capitalization. Our model outperforms state-of-the-art benchmarks on most measures and significantly improves the Sharpe ratio. It continues to outperform baselines even after accounting for the high transaction costs associated with trading cryptocurrencies.
\end{abstract}


\maketitle

\section{Introduction}
Cross-sectional strategies are a popular class of trading strategies that involve first sorting assets by some ranking criteria and then trading the winners against losers.
Since the seminal work of \cite{jegadeeshReturnsBuyingWinners1993} documenting the strategy and its performance trading US equities, numerous technical refinements have been introduced. For instance, the ranking step, which is at the core of the model, evolved from being performed with heuristics \cite{jegadeeshReturnsBuyingWinners1993} to involving more sophisticated DL (Deep Learning) prediction models \cite{guAutoencoderAssetPricing2019, kimEnhancingMomentumStrategy2019} and modern ranking algorithms such as LambdaMART \cite{burgesRankNetLambdaRankLambdaMART2010, pohBuildingCrossSectionalSystematic2021, wangStockRankingMarket2018}. 

Fundamental to calibrating these models is the availability of sufficient data. While this is somewhat mitigated for DL-based strategies operating in the high-frequency domain or for mature assets with longer trading histories, deploying these strategies on newer instruments is challenging due to their small sample sizes, and their degraded performance will likely result in substantial trading losses. Cryptocurrencies, which have recently seen broad interest in academia and industry, are an example of such a class of instruments.

This dilemma reflects the broader challenge of training DL models that generalise well when confronted with limited data. Although numerous approaches (reducing model complexity \cite{brigatoCloseLookDeep2020a}, few-shot learning \cite{wangGeneralizingFewExamples2021} and data augmentation techniques \cite{fengSurveyDataAugmentation2021, shortenSurveyImageData2019}) exist for tackling this issue, Transfer Learning stands out as a promising methodology.
Transfer Learning focuses on transferring the knowledge obtained across different but related source or upstream domains in order to boost the performance of a decision-making function on some target or downstream domains (we will use `source' and `upstream' interchangeably, the same applies to `target' and `downstream') of interest \cite{zhuangComprehensiveSurveyTransfer2019, daySurveyHeterogeneousTransfer2017}. It is a broad and active research area, and given the nuances that different problems entail, it overlaps several fields such as domain adaptation and multi-task learning\footnote{Domain adaptation is used when the source and target datasets belong in different domains \cite{wangGeneralizingUnseenDomains2021}. Multi-task learning is used when the goal is to jointly learn a group of related tasks by exploiting the interconnections between tasks \cite{zhuangComprehensiveSurveyTransfer2019}.}.
The multitude of transfer learning successes in computer vision \cite{kolesnikovBigTransferBiT2020, shenPartialBetterAll2021} and NLP (natural language processing) \cite{devlinBERTPretrainingDeep2019} tasks have boosted its popularity both as a go-to solution for knowledge transfer problems, as well as a research topic in its own right.

In the space of IR (information retrieval), transfer learning and its closely associated fields have been documented to significantly improve the accuracy of sorting algorithms in target domains with limited samples \cite{chenKnowledgeTransferCross2010, chapelleMultitaskLearningBoosting2010a, gaoLearningRankOnly2010, liSelflabelingMethodsUnsupervised2020}. However, unlike the computer vision and NLP communities, which have rapidly adopted transformer-based solutions in response to their domain-specific transfer learning problems, we note that a corresponding development in IR is surprisingly absent. Inspired by recent works of \cite{pobrotynContextAwareLearningRank2020} and \cite{peiPersonalizedRerankingRecommendation2019} which exploit the Transformer's attention module to improve ranking performance as well as a first adaptation in the context of finance \cite{pohEnhancingCrossSectionalCurrency2022}, we similarly make use of this mechanism and propose the Fused Encoder Networks (FEN). Central to the model's architecture is the usage of dual encoder blocks -- one of which is optimised on a larger but related upstream dataset and then deployed in conjunction with the other encoder block calibrated on a smaller downstream dataset of interest. We focus on applying our transfer ranking model to cryptocurrencies, where relatively short trading histories (relative to more mature financial instruments) and recent institutional interest promotes them as a practical use-case. 
Applying cross-sectional momentum across ten cryptocurrencies with the largest market capitalisation, we provide evidence of the superiority of our approach over a set of benchmarks which includes state-of-the-art LTR (Learning to Rank) algorithms. We obtain a threefold boost in the Sharpe ratio over risk-adjusted momentum and a gain of around 50\% over the best baseline model. 
We structure the rest of the paper as follows: The next section discusses related works applying transfer learning to finance and ranking tasks. We next move on to sections that define the problem and describe the proposed model's key elements. We subsequently evaluate the results of our study and finally conclude this work.

\section{Related Works}
\subsection{Transfer Learning in Finance}\label{sec:lit_transfer_learning_in_finance}

Transfer learning focuses on producing an effective model for a target task with scarce training samples by transferring knowledge across different but related source domain(s) \cite{zhuangComprehensiveSurveyTransfer2019, daySurveyHeterogeneousTransfer2017}. It is an active area of study in machine learning, enjoying success and widespread adoption in various applications \cite{doTransferLearningText2005, daiCoclusteringBasedClassification2007, kornblithBetterImageNetModels2019, blumbergMultistagePredictionNetworks2019}. Given the richness of the literature, numerous categorisation schemes have been established to distinguish the diverse yet interconnected strands of transfer learning research. For instance, the work in \cite{panSurveyTransferLearning2010} states that depending on the availability of label information, problems in transfer learning can be categorised as transductive, inductive, and unsupervised\footnote{Inductive transfer learning refers to the scenario where labelled data is available for both the source and target domains. The labelled source data is available in transductive transfer learning, but the labelled target domain data is not. Finally, in unsupervised transfer learning, the label information is unavailable for both source and target domains.}. Alternatively, depending on the source and target domains represented in similar feature or label spaces, transfer learning can be loosely classed as either homogeneous or heterogeneous \cite{zhuangComprehensiveSurveyTransfer2019}. Using the schema in \cite{panSurveyTransferLearning2010}, transfer learning techniques divides into four groups: instance-based, feature-based, parameter-based, and relational-based approaches\footnote{Instance-based techniques are primarily based on the instance weighting strategy, while feature-based approaches construct a new representation from the original set of features. Parameter-based methods typically involve transferring knowledge at the model or parameter level. Lastly, relational-based techniques generally focus on relational problems and extract the rules or the logical relationship from the source for use in the target domain.}. Despite these attempts to provide a comprehensive taxonomy of the research landscape, several terminology inconsistencies persist -- for example, phrases such as domain adaptation and transfer learning are used interchangeably to refer to the same processes \cite{weissSurveyTransferLearning2016}. We point our readers to  \cite{panSurveyTransferLearning2010}, \cite{weissSurveyTransferLearning2016}, and \cite{zhuangComprehensiveSurveyTransfer2019} for a better understanding of this broad field.

In the context of transfer learning for finance, most works focus on trading and employ an inductive approach with a parameter-based focus -- incorporating the source architecture's weights or model components into the target model to optimise learning the task of interest. 
In one of the earliest examples of transfer learning in finance, \cite{zhangDeepLOBDeepConvolutional2019} propose DeepLOB, a hybrid deep neural model that predicts stock price movements using high-frequency limit order data. The authors document DeepLOB's ability to extract universal features by showing that it can successfully predict the direction of price movement of a target set of stocks that are not part of the source training set. \cite{nakagawaRobustTransferableDeep2020} develops the RIC-NN (Rank Information Coefficient Neural Net) -- a multi-factor approach for stock price prediction using deep transfer learning. They demonstrate that the lower layers of the optimised model's weights in one region (represented by MSCI North America) can initialise the same model for prediction in a different region (MSCI Asia Pacific). 
In a similar vein but taking on a broader coverage, \cite{koshiyamaQuantNetTransferringLearning2021} propose QuantNet, which fuses information by training on different markets with a global bottleneck shared across multiple autoencoder-based architectures. The model then employs a separate decoder for each target market to generate market-specific trading strategies. 
The work in \cite{jeongImprovingFinancialTrading2019} develops a deep Q-learning reinforcement learning trading agent that they combine with pre-trained weights to prevent over-fitting from insufficient financial data. Concentrating on financial time-series forecasting, \cite{heTransferLearningFinancial2019} tackles the poor performance of training deep learning models on limited data with multi-source transfer learning. In a risk-management setting but utilising a similar pre-trained weights approach, \cite{suryantoTransferLearningCredit2019} investigate the utility of knowledge transfer methods in evaluating credit risk.

Unlike earlier studies applying transfer learning to the problems in finance, ours differ in a few key aspects.
Firstly, \cite{koshiyamaQuantNetTransferringLearning2021} and \cite{nakagawaRobustTransferableDeep2020} primarily concentrate on transferring knowledge and do not address model overfitting. 
Additionally, while \cite{jeongImprovingFinancialTrading2019} and \cite{heTransferLearningFinancial2019} are closer in spirit to our work (i.e., they aim to alleviate overfitting that surfaces when calibrating on the target task with limited samples), both do not explicitly connect their research to momentum effects. 

Although momentum effects have been documented on numerous asset classes \cite{asnessValueMomentumEverywhere2013, menkhoffCurrencyMomentumStrategies2012, pirrongMomentumFuturesMarkets2005,luuMomentumGovernmentBondMarkets2012}, and with numerous sophisticated prediction models developed to enhance the profitability of momentum-based strategies \cite{limEnhancingTimeSeries2019, kimEnhancingMomentumStrategy2019, woodSlowMomentumFast2022}, the literature on pre-training these advanced models and applying them to settings where the dataset is small, is scarce. 
With cryptocurrencies, numerous studies (e.g., \cite{BELLOCCA2022100310} and \cite{guoTimeVaryingNetworkCryptocurrencies2022}) apply sophisticated machine learning for momentum-based trading. However, few, if any, explore augmenting these trading models by transferring knowledge from different datasets and examining their application over lower frequency settings, such as over weeks (which exacerbates the data scarcity problem). We address this deficiency in the research literature with our work in this paper.

\subsection{Transfer Ranking}\label{sec:lit_transfer_ranking}
Transfer Ranking (TR) is defined as the application of transfer learning to LTR algorithms \cite{liTransferLearningInformation2019, liuTransferRanking2011}. While a plethora of TR algorithms have been developed to address the extensive nature of ranking tasks \cite{chapelleFutureDirectionsLearning2011}, the primary objectives of these algorithms are essentially similar, i.e., to boost the ranking model's accuracy on a new collection of items (which we refer to as the target or downstream dataset) by leveraging labelled data from another collection (source or upstream dataset) \cite{liTransferLearningInformation2019}. From the literature classification in \cite{liTransferLearningInformation2019},
TR techniques are either supervised or unsupervised, with each category further divided into homogeneous or heterogeneous depending on the similarity between the source and target feature spaces. Given that our work resembles model adaptation within the space of homogeneous supervised TR techniques, we review works related to this form and point the reader to \cite{liTransferLearningInformation2019} for details on other setups. 

Model adaptation TR concentrates on adapting the existing source model with a few labelled training data from the target collection. Most algorithms concentrating on this form treat the source ranker as a base model or exploit it as a feature for training \cite{liTransferLearningInformation2019}. These methods respectively resemble the parameter- and feature-based transfer learning methods in the scheme proposed by \cite{panSurveyTransferLearning2010}.
The work in \cite{gaoModelAdaptationModel2009}, for instance, explores two classes of model adaptation techniques for web search ranking: (i) model interpolation and (ii) error-driven learning approaches based on boosting. The first method learns an ensemble involving both source and target models; In the second, the authors propose LambdaBoost and  LambdaSMART \cite{burgesRankingBoostingModel2008}, which at each boosting iteration adapt the source model to target data (the former searches the space of finite basis functions, while the latter searches the space of regression trees).
Other research directions involve calibrating the source model to better align with the target data. For example, \cite{chenTradaTreeBased2008} proposes Trada, which is a tree-based model that tunes both weights and splits based on a combination of source and target data. Trada is further improved by incorporating pairwise information to refine the model's adaptation process \cite{baiCrossmarketModelAdaptation2010}. Lastly, \cite{wangPersonalizedRankingModel2013} adopts a different approach by coefficient transformation. Concretely, the authors develop an algorithm that generates personalised results by learning a transformation matrix that manipulates the weights of a global ranker. 

Surveying the literature surrounding model adaptation TR, solutions employing trees are common. This development is likely owing to their flexibility and robustness, which has been validated by the performant tree-boosting ranking algorithm, LambdaMART \cite{burgesRankNetLambdaRankLambdaMART2010}. 
While recent LTR breakthroughs, such as \cite{qinAreNeuralRankers2021} and \cite{pobrotynContextAwareLearningRank2020} provide clear evidence that transformer-based rankers can outperform their tree-based counterparts, we note a lack of models employing this architecture in TR. To this end, we propose the Fused Encoder Networks (FEN) -- a novel and hybrid parameter-sharing transfer ranking model inspired by the Transformer. The FEN fuses and leverages information flexibly extracted by a pair of encoder-attention modules operating on both source and target domain data. With this design, the model inherits the advantages of self-attention and can account for dependencies across instruments at the loss level during training and inference time. By incorporating the components of the source model in the FEN model, the networks then determine the optimal combination of source and target information for the ranking task.

\section{Problem Definition} \label{sec:problem_definition}

This section formalises the problem setting. To get a better idea of the LTR framework as applied to cross-sectional momentum and how TR builds on top of LTR, please refer to the section titled \nameref{app:transfer_ranking_framework} in the Appendix.

\subsection{Linking Returns and Model Formulation}

Given a portfolio of cryptocurrencies that is rebalanced weekly, the returns for a cross-sectional momentum (CSM) strategy at week $t$ can be expressed as follows:
\begin{align}
r^{CSM}_{t, t+1} = \frac{1}{n} \sum^{n}_{i=1} S^{(i)}_t \left ( \frac{\sigma_{tgt}}{\sigma^{(i)}_t} \right ) r^{(i)}_{t, t+1}
\label{eqn:csm_rets} 
\end{align}
where $r^{CSM}_{t,t+1}$ refers to the portfolio's realised returns over weeks $t$ to $t+1$, $n$ denotes the number of cryptocurrencies and $S^{(i)}_t \in \{-1, 0, 1\}$ characterises the CSM signal or trading rule for instrument $i$.
We fix the annualised volatility target $\sigma_{tgt}$ at 15\% and scale returns with $\sigma^{(i)}_t$ which is an estimator for ex-ante weekly volatility. The latter is computed as the rolling exponentially weighted standard deviation with a 26-week (approximately half a year) span on weekly returns. 

The CSM signal $S_i$ in Equation (\ref{eqn:csm_rets}) can be obtained using heuristics \citep{jegadeeshReturnsBuyingWinners1993} or by minimising the empirical loss of some prediction model $f$ parameterised by $\theta$:
\begin{equation}
    \mathcal{L}(\theta) = \frac{1}{|\Psi|} \sum_{(\textbf{x}, \textbf{y}) \in \Psi} \ell 
    \bigl (\textbf{y}, f(\textbf{x}; \theta) \bigr ) \label{eqn:emp_loss}
\end{equation}
With slight abuse of notation, we take each item $x_i$ to be a $k$-sized real-valued feature vector. Since $\textbf{x}$ is a list of $n$ items, we have $\textbf{x}\in\mathbb{R}^{n \times k}$.
The training data is denoted as $\Psi=\{(\textbf{x}, \textbf{y}) \in \mathbb{R}^{n \times k} \times \mathbb{Z}_{+}^{n}\}$, where $\textbf{y}$ is a list of $n$ bins for the same $\textbf{x}$ sorted by realised returns for the \textit{next} period. When the prediction model $f$ adopts a regress-then-rank approach such as those proposed in \citet{guAutoencoderAssetPricing2019} and \citet{kimEnhancingMomentumStrategy2019}, then the loss function $\ell$ is usually the pointwise MSE (mean squared error). With LTR models, pairwise and listwise losses that account for the mutual interaction among objects are usually used for $\ell$.

\subsection{Model Formulation under the Transfer Ranking Framework}
With parameter-based transfer learning approaches, an upstream or source model is first trained on the source data. This process can be expressed as:
\begin{equation}
    \mathcal{L}(\theta_S) = \frac{1}{|\Psi_S|} \sum_{(\textbf{x}, \textbf{y}) \in \Psi_S} \ell 
    \bigl (\textbf{y}, f_S(\textbf{x}; \theta_S) \bigr ) \label{eqn:parameter_sharing_source_loss}
\end{equation}

\begin{figure*}[t!]
  \centering
  \includegraphics[width=\linewidth]{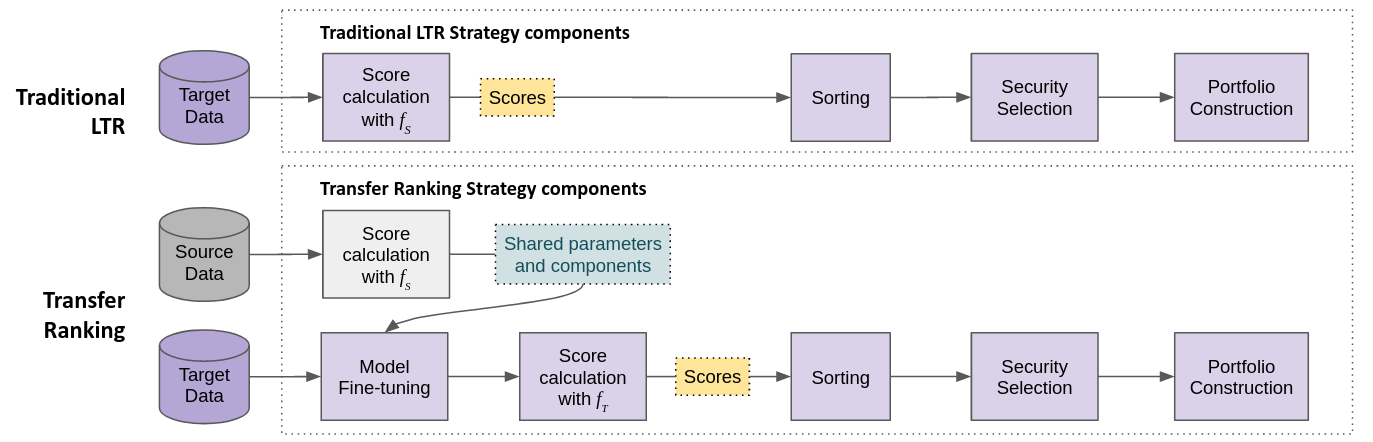}
  \caption{Pipelines for traditional LTR (top) and Transfer Ranking (bottom): The latter involves first training a source (upstream) model on a different but related dataset and then transferring the knowledge by sharing the calibrated weights/components with a target (downstream) model. Apart from requiring an additional fine-tuning step, the training process for the target model follows that of traditional LTR.}\label{fig:pipeline}
\end{figure*}
\begin{figure*}[t!]
  \centering
  \includegraphics[width=\linewidth]{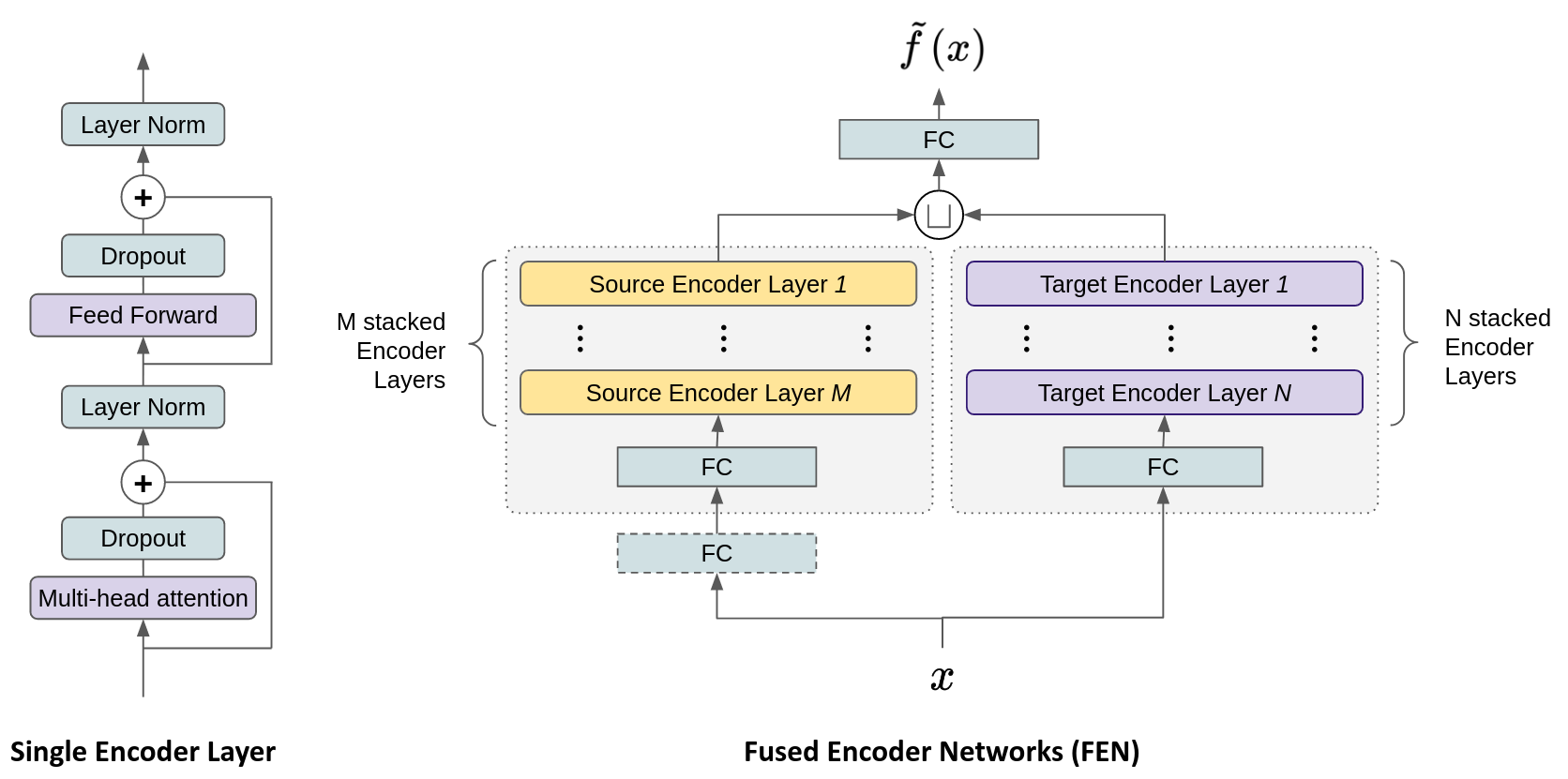}
  \caption{Architecture of a single Encoder Layer (left) as well as the FEN (Fused Encoder Networks) (right). Each single Encoder Layer comprises a multi-head attention module and a feed-forward network and uses dropouts and layer normalisation to learn higher-order feature representations. The FEN comprises both the source and target encoder blocks running in parallel. FC and the $\sqcup$ symbol represents a fully connected layer and the concatenate operator respectively. The FC block with the dashed outline is optional and is used when the target and source inputs differ in dimensions.}\label{fig:schematic}
\end{figure*}

where the terms $f_S, \theta_S \textrm{ and } \Psi_S$ are respectively the prediction model, its parameter vector, and the associated dataset. The tuned upstream weights are then incorporated into the downstream or target model:
\begin{equation}
    \mathcal{L}(\theta_T) = \frac{1}{|\Psi_T|} \sum_{(\textbf{x}, \textbf{y}) \in \Psi_T} \ell 
    \bigl (\textbf{y}, f_T(\textbf{x}; \theta_S, \theta_T) \bigr ) \label{eqn:parameter_sharing_target_loss}
\end{equation}

where adhering to the form laid out by Equation (\ref{eqn:emp_loss}), $f_T$ refers to the ranking function parameterised by both source and target parameters $\theta_S \textrm{ as well as } \theta_T$, and $\ell$ is either a pairwise or listwise loss. In the final fine-tuning step, minimising the loss can be formulated as:
\begin{align}
    \mathcal{L}(\Theta_S, \theta_T) = \frac{1}{|\Psi_T|} \sum_{(\textbf{x}, \textbf{y}) \in \Psi_T} \ell 
    \bigl (\textbf{y}, f_T(\textbf{x}; \Theta_S, \theta_T) \bigr ) \label{eqn:hybrid_parameter_sharing_loss}
\end{align}
and $\Theta_S$ denotes the subset of parameters within $\theta_S$ that is permitted to be fine-tuned. Using this final model, we compute ranking scores which we then sort and threshold to obtain the cross-sectional trading signal $S_t$ required for portfolio construction. Figure \ref{fig:pipeline} visualises the TR pipeline and compares it to traditional LTR.

\section{Fused Encoder Networks}\label{sec:fen}
Optimising the FEN (Fused Encoder Networks) involves training a source model on a larger dataset. The calibrated components are subsequently shared with the target model (i.e., FEN) as part of a broader framework. The following section provides vital details and explains how various components come together.

\subsection{Source Model Architecture} 
We use a stacked encoder block with a linear projection applied to inputs (similar to those used in \citet{pobrotynContextAwareLearningRank2020} and \citet{pohBuildingCrossSectionalSystematic2021}) as our source model. Each encoder within the stack is a composite layer housing the attention mechanism and a feed-forward network that enables it to learn higher-order feature representations. It is an essential component in both the source and target models (See left of Figure \ref{fig:schematic}).

\subsection{Feed-forward Network} This component introduces non-linearity via its ELU (Exponential Linear Unit) \citep{clevertFastAccurateDeep2016} activation. It also facilitates interactions across different parts of its inputs.

\subsection{Self-Attention and Multi-head Attention}
The attention mechanism is a function that maps a query and a key-value pair to an output. While a few variants of attention exist, we use the form similar to that employed in the Transformer, which is the scaled dot-product attention followed by a softmax operation \citep{vaswaniAttentionAllYou2017}:
\begin{equation}
    \textrm{Att}(\textbf{Q},\textbf{K},\textbf{V}) = \textrm{softmax}\Biggl ( \frac{\textbf{QK}^\top}{\sqrt{d_{model}}} \Biggr ) \textbf{V} \label{eqn:attention}
\end{equation}
where \textbf{Q}, \textbf{K} and \textbf{V} are respectively the query, key and value matrices. In \textit{self}-attention, the \textit{same} inputs are used for all three matrices.
Combining multiple attention units produces the multi-head attention module, which boosts the model's capacity to learn representations:
\begin{align}
    \textrm{MHA}(\textbf{Q},\textbf{K},\textbf{V}) & = \textrm{concat}(\textrm{head}_1, ... , \textrm{head}_h)\textbf{W}^O \\
    \textrm{head}_i & = \textrm{Att}(\textbf{QW}_i^\textbf{Q}, \textbf{KW}_i^\textbf{K}, \textbf{VW}_i^\textbf{V})
\end{align}
where in the above equations, each $\textrm{head}_i$ (out of $h$ heads) refers to the $i$th attention mechanism of Equation (\ref{eqn:attention}), and learned parameter matrices:
\begin{align}
  & \textbf{W}_i^\textbf{Q} \in \mathbb{R}^{{d_{model}}\times d_q}, \quad
  \textbf{W}_i^\textbf{K} \in \mathbb{R}^{{d_{model}}\times d_k}, \nonumber \\
  & \textbf{W}_i^\textbf{V} \in \mathbb{R}^{{d_{model}}\times d_v}, \quad
  \textbf{W}^\textbf{O} \in \mathbb{R}^{hd_v\times d_{model}}  \nonumber
\end{align}
where typically $d_q=d_k=d_v=d_{model}/h$ is used.

\subsection{Stacking Encoder Layers} 
A single encoder block $\xi(\cdot)$ can be expressed as:
\begin{align}
    \xi(\textbf{x}) & = \Gamma(\textbf{z} + \delta(\phi(\textbf{z}))) \\
    \textbf{z} & = \Gamma(\textbf{x} + \delta(\textrm{MHA}(\textbf{x})))
\end{align}
where $\Gamma(\cdot), \delta(\cdot) \textrm{ and } \phi(\cdot)$ refers to the layer normalisation operation, a dropout function, and a projection onto a fully-connected layer respectively. Additionally, MHA$(\cdot)$ represents the multi-head attention module. Stacking multiple encoder layers enables the model to learn more complex representations:
\begin{equation}
    \boldsymbol{\xi}(\textbf{x}) = \xi_1 \bigl(...(\xi_N(\textbf{x})) \bigr )\label{eqn:enc_stack}
\end{equation}
where $\boldsymbol{\xi}(\textbf{x})$ in Equation (\ref{eqn:enc_stack}) characterises a stacked encoder block involving a series of $N$ encoder layers.

\subsection{Target Model Architecture}
With conventional parameter-sharing techniques, if a neural model is used to learn the source task, then a target model can be constructed by directly retaining most of its layers
 \citep{zhuangComprehensiveSurveyTransfer2019}. 
Motivated by the empirical superiority of context-aware LTR models using the attention mechanism over standard rankers \citep{pobrotynContextAwareLearningRank2020, pohEnhancingCrossSectionalCurrency2022}, we utilise a hybrid approach -- running the pre-trained source Transformer's encoder block $\boldsymbol{\xi}_S(\cdot)$ as an additional feature extractor operating in \textit{parallel} with the target Transformer's block $\boldsymbol{\xi}_T(\cdot)$:
\begin{align}
    \boldsymbol{\xi}_S(\textbf{x}) = \xi_{S_1} \bigl(...(\xi_{S_M}(\textbf{x})) \bigr )\label{eqn:enc_stack_source} \\
    \boldsymbol{\xi}_T(\textbf{x}) = \xi_{T_1} \bigl(...(\xi_{T_N}(\textbf{x})) \bigr )\label{eqn:enc_stack_target}
\end{align}
where $\xi_{S_i}$ and $\xi_{T_j}$ are respectively the $i$th and $j$th encoder layer from the source and target model (See the right of Figure \ref{fig:schematic}). 
Note the additional full-connected block with the dashed outline -- this is optional and is used when the target and source inputs differ in dimensions.
Against standard parameter-sharing methods, our setup has the additional advantage of being plug-and-play, allowing to be appended anywhere to any pre-trained network. The outputs from both encoder blocks are concatenated:
\begin{align}
    \boldsymbol{\xi}_{S,T}(\textbf{x}) & = \textrm{concat}\bigl (\boldsymbol{\xi}_S(\textbf{x}),
    \boldsymbol{\xi}_T(\textbf{x}) \bigr)
\end{align} 
before projecting onto a fully connected layer to obtain our raw score predictions $f(\textbf{x})$. Although we use a simple projection for simplicity, this can be replaced by more complex processing such as those that involve deep or cross networks \citep{wangDCNV2Improved2021}.

\subsection{Loss Function}
We optimise both the source and target components with the ListNet loss \citep{caoLearningRankPairwise2007} and note that other losses such as ListMLE \citep{xiaListwiseApproachLearning2008} or metrics like ApproxNDCG \citep{bruchRevisitingApproximateMetric2019} can be used. The ListNet loss is defined as:
\begin{equation}
    \ell(\textbf{y}, f(\textbf{x})) = -\sum_i \textrm{softmax}(\textbf{y})_i \times \log \bigl (\textrm{softmax}(f(\textbf{x}))_i \bigr )
    \label{eqn:listnet_loss}
\end{equation}

\subsection{Target Model Fine-tuning}

Deep networks can be fine-tuned in several ways. A straightforward approach is to optimise all the parameters of the model after initialising it with the pre-trained model's parameters. Another method, driven by the findings that learned features transition from general to specific along a network \citep{longLearningTransferableFeatures2015}, concentrates on re-training only the last few layers while leaving the parameters of the remaining initial layers frozen (at their pre-trained values). 
There are also other methods, such as those proposed by \citet{girshickRichFeatureHierarchies2014} and \citet{razavianCNNFeaturesOfftheShelf2014}. As the number of layers to freeze is a manual design choice that introduces unnecessary complexity, we do not freeze any layers and instead focus on re-training the entire model with a low learning rate.

\section{Performance Evaluation Details}

\subsection{Dataset Overview}

For actual performance evaluation, all models utilise daily closing prices obtained from CoinMarketCap  \citep{coinmarketcap_cryptos} further downsampled to the weekly frequency. 
Our universe is defined as the top 10 cryptocurrencies by market capitalisation as of the end of Dec-2019, with at least four years of price history. Given the limited size of the data, this selection ensures sufficient samples for model training and drawing meaningful inferences. We fix this set of cryptocurrencies from the end of Dec-2019 to the end of Dec-2021 to minimise the impact of survivorship bias. The complete data set thus runs from 1-Jan-2016 to 31-Dec-2021, with 2018 and 2021 being the first and last target test years respectively. 

To train the source model needed for knowledge transfer, we make use of the same set of daily data relating to 30 currency pairs \footnote{Indeed, it can be argued that other source data sets are more suitable for pre-training given the objective of this study. Returns on technology stocks might be one such example. We leave this for future research.}
as per \citet{bazDissectingInvestmentStrategies2015} and \citet{pohEnhancingCrossSectionalCurrency2022} obtained from the Bank for International Settlements (BIS) \citep{bis_currencies} spanning May-2000 to Dec-2021 which we again downsample to the weekly frequency. To measure risk aversion, we use the daily close of the VIX historical data from the Cboe Global Markets \citep{vix_historical_data}, where a week is labelled risk-off if it contains one or more days when the VIX is 5\% higher than its 60-day moving average. A normal week is thus a non-risk-off week.
For further details about the data, we refer the reader to \nameref{app:data} in the Appendix. 

\subsection{Strategy and Predictor Descriptions}
Our portfolios are rebalanced weekly. We construct an equally weighted long/short portfolio involving the top/bottom two cryptocurrencies as ranked by the model and calculate returns using Equation (\ref{eqn:csm_rets}). 

We note that there are also other approaches for constructing the cross-sectional portfolio. Within the cryptocurrency literature, our method resembles \citet{tzouvanasMomentumTradingCryptocurrencies2020}, who forms a self-financed long/short strategy based on a risk-adjusted momentum portfolio. Given that volatility-scaling is a common feature in time-series momentum \citep{moskowitzTimeSeriesMomentum2012, baltasMomentumStrategiesFutures2012}, it is plausible that our approach captures similar effects to this class of strategies. \citet{liuCommonRiskFactors2022} adopt a double-sort approach -- first sorting based on market capitalisation, and then by three-week momentum. The final portfolios are located at the intersection of this two-step process and are traded without volatility-scaling. This approach resembles the momentum factor portfolio of \citet{famaChoosingFactors2018} in the asset pricing research. Finally, in the equity literature, there is \citet{jegadeeshReturnsBuyingWinners1993} who trade equally-weighted winner and loser decile portfolios with no risk scaling. 

We use a simple set of returns-based features for predictors across both source and target models. The source model is trained with the following features:
\begin{enumerate}
    \item \textit{Raw returns} -- Raw FX returns over the past 1, 2, 3, and 4 weeks, using the findings of \citet{menkhoffCurrencyMomentumStrategies2012}. 
    \item \textit{Normalised returns} -- Raw FX returns over the previous 1, 2, 3, and 4 weeks standardised by weekly volatility scaled to the appropriate time scale.
\end{enumerate}
For the target model, we again use both raw and normalised cryptocurrency returns but only over the past 1, 2, and 3 weeks as evidence of significant momentum returns in cryptocurrencies is mixed using a 1-month formation period
\citep{tzouvanasMomentumTradingCryptocurrencies2020}.?? see www.packhacker.com

\subsection{Models and Comparison Metrics} \label{sec:benchmarks_metrics}
\begin{enumerate}
    \item 1-week Returns (1WR) -- This is risk-adjusted momentum strategy based on the findings of \citet{tzouvanasMomentumTradingCryptocurrencies2020}. Unlike the original approach, we do not form overlapping portfolios but instead construct weekly long/short portfolios based on returns over the previous week. The final portfolio is scaled to the desired target volatility.
    \item Multi-Layer Perceptron (MLP) -- Characterises the typical Regress-then-rank model used by contemporary strategies. 
    \item ListNet (LN) -- Listwise Learning to Rank model proposed by \citet{caoLearningRankPairwise2007}. 
    \item LambdaMART (LM) -- Pairwise Learning to Rank model proposed by \citet{burgesRankNetLambdaRankLambdaMART2010}. This model is based on Gradient Boosted Decision Trees (GBDT) and is still regarded as the state-of-the-art for LTR problems in recent IR studies \citep{qinAreNeuralRankers2021}.
    \item Self-attention-based Ranker (SAR) -- This benchmark uses the Transformer's encoder module for ranking and is applied directly to the cryptocurrency data. The use of the encoder as a scoring function for LTR was proposed by \citet{pobrotynContextAwareLearningRank2020}. With this approach, they establish state-of-the-art performance on the WEB30K LTR dataset\footnote{WEB30K or MSLR WEB30K (Microsoft Learning to Rank Datasets-30k) is a large scale dataset released by Microsoft that is used for researching and benchmarking LTR algorithms.} and report substantial performance gains over the well-known and high performing LambdaMART. 
    This idea was later applied to constructing cross-sectional strategies by \citet{pohEnhancingCrossSectionalCurrency2022}\footnote{Unlike \citet{pohEnhancingCrossSectionalCurrency2022}, we use the variant without positional encodings as we are not performing re-ranking.}.
    \item SAR with Parameter-Sharing (SAR+ps) -- This involves pre-training the Transformer's encoder module on the larger FX dataset and then fine-tuning on the target cryptocurrency set. Fine-tuning pre-trained models has been documented to be a valid and successful transfer learning approach for computer vision \citep{zhuangComprehensiveSurveyTransfer2019} and NLP \citep{devlinBERTPretrainingDeep2019, howardUniversalLanguageModel2018} tasks.
    \item Fused Encoder Networks (FEN) -- The proposed model.
\end{enumerate}

We assess the models over four key areas (profitability, riskiness, risk-adjusted performance, and ranking accuracy) with the following metrics:

\begin{itemize}
    \item Profitability: Expected returns ($\mathbb{E}[\textrm{Returns}]$) and Hit rate (percentage of positive returns at the portfolio level obtained over the out-of-sample period).
    \item Risks: Volatility, Maximum Drawdown (MDD), and Downside Deviation.
    \item Financial Performance: Sharpe $\Bigl(\frac{\mathbb{E}[\textrm{Returns}]}{\textrm{Volatility}}\Bigr)$, Sortino $\Bigl(\frac{\mathbb{E}[\textrm{Returns}]}{\textrm{MDD}}\Bigr)$ and Calmar $\Bigl(\frac{\mathbb{E}[\textrm{Returns}]}{\textrm{Downside Deviation}}\Bigr)$ ratios are used as a gauge to measure risk-adjusted performance. We also include the average profit divided by the average loss $\Bigl(\frac{\textrm{Avg. Profits}}{\textrm{Avg. Loss}}\Bigr)$.
    \item Ranking Performance: NDCG@2. This is based on the Normalised Discounted Cumulative Gain (NDCG) \citep{jarvelinIREvaluationMethods2000}, a measure of graded relevance. We focus on assessing the NDCG of the top/bottom two cryptocurrencies, which are linked directly to strategy performance.
\end{itemize}

\subsection{Training and Backtest Details}

\textit{Benchmark Models:}

\begin{figure*}[]
  \centering
  \includegraphics[scale=0.5]{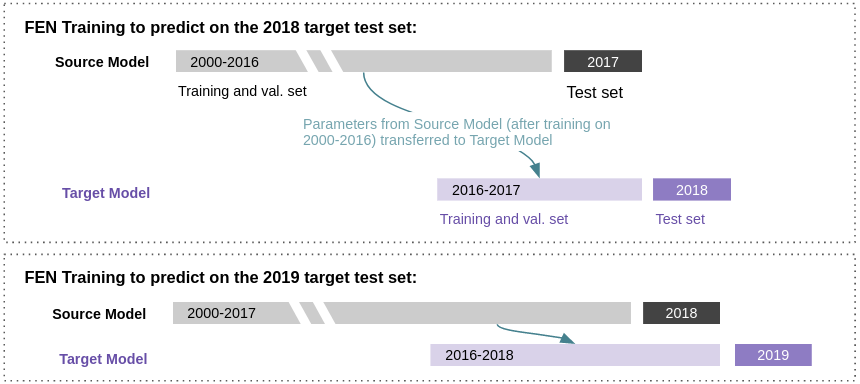}
  \caption{Source and target (FEN) model training. The source and target models are trained using an expanding window of one-year blocks. To minimise any impact of negative transfer, we assess the performance of the upstream model in the year preceding the target test year. We only use models that have performed well for knowledge transfer.
  }\label{fig:fen_training}
\end{figure*}

Except for 1WR and LM, all models are tuned with minibatch stochastic gradient descent with the \texttt{Adam} optimiser.
All machine-learning-based models are calibrated using an expanding window approach in one-year blocks. Using the first iteration as an example: we retain 90\% of the 2016-2017 period for training and use the remainder for cross-validation, with 2018 held out for testing. In the next iteration, the training set is expanded to include 2018 (i.e., 2016 to 2018), and the model is tested using 2019 (See also Figure \ref{fig:fen_training}).
Backpropagation was conducted for a maximum of 100 epochs and with an expanding window approach, where 90\% of a given training block is used for training and the remainder for validation.
Early stopping was used to prevent over-fitting and is triggered when the validation loss does not improve for 25 epochs. Model hyperparameters were tuned over 50 iterations of search using \texttt{HyperOpt} \citep{bergstraHyperoptPythonLibrary2015}. 

Further details relating to model calibration can be found in \nameref{app:training_details} of the Appendix.

\textit{Fused Encoder Networks:}

The FEN is optimised in three steps. 
The upstream model is first calibrated on the source FX data. This is performed with a one-year expanding window approach where we test our strategy on a given year and use all preceding samples for model tuning with a 90\%/10\% train/validation split. For the next year, we expand and shift the training and test periods respectively. 

For the next step, the source model's encoder block is transferred to the FEN and trained on the target data with the source block frozen. Like source model training, the FEN is trained using a single-year expanding window with a 90\%/10\% split. 
To minimise any detrimental effects arising from \textit{negative transfer}, which is the case when the transferred source knowledge adversely impacts performance on the downstream task, we first proceed as follows: to generate predictions for each out-of-sample year $t$ on the target task, we produce ten runs of the source models evaluated on $t-1$. We pick the source model with the \textit{highest} Sharpe ratio and use it to train the target model (as we evaluate model performance (discussed later) based on ten runs, we pick the top 2 source models and go on to train five target models with each.).
This process is illustrated in Figure \ref{fig:fen_training}. The consequences of transferring knowledge from models with the \textit{lowest} Sharpe ratios can be found in \nameref{app:negative_transfer} in the Appendix.

The final fine-tuning phase optimises the unfrozen network with a very low learning rate -- set as the smallest value in the search space for learning rates. The model is run for a maximum of 100 epochs and is stopped early if the validation loss fails to improve for ten epochs. We fine-tune only the FEN and SAR+ps. Further details can be found in \nameref{app:training_details} in the Appendix.

\section{Results and Discussion}
\subsection{Trading Performance} \label{sec:trading_performance}

\begin{table*}[t!]\centering
\caption{Performance metrics for all models over the out-of-sample 2018-2021 period, rescaled to match a 15\% annual volatility target. All figures except those belonging to 1WR are computed based on ten runs. FEN outperforms the benchmark models on nearly all measures.}
\label{tbl:trading_performance_comparison}
\begin{tabular}{@{}rccccccccc@{}}
\toprule
& \multicolumn{6}{c}{\textbf{Benchmarks}}& \phantom{abc}& \multicolumn{2}{c}{\textbf{Proposed}} \\
\cmidrule{2-7}\cmidrule{9-10} & 1WR & MLP & LN & LM & SAR & SAR+ps && FEN \\
\midrule
E[returns] 	            & 0.053 & 0.080$\pm$0.043 & -0.094$\pm$0.132 & 0.076$\pm$0.109 & 0.084$\pm$0.100 & -0.006$\pm$0.074 && \textbf{*0.141$\pm$0.080} & \\
Volatility              & 0.221 & 0.201$\pm$0.033 &  0.211$\pm$0.087 & 0.205$\pm$0.068 & \textbf{*0.176$\pm$0.015} &  0.178$\pm$0.012 && 0.187$\pm$0.012 & \\
Sharpe Ratio            & 0.238 & 0.403$\pm$0.240 & -0.397$\pm$0.539 & 0.456$\pm$0.546 & 0.496$\pm$0.583 & -0.035$\pm$0.406 && \textbf{*0.749$\pm$0.410} & \\
Sortino Ratio           & 0.484 & 0.692$\pm$0.436 & -0.405$\pm$0.745 & 0.943$\pm$1.061 & 0.937$\pm$1.070 &  0.028$\pm$0.597 && \textbf{*1.526$\pm$1.023} & \\
Calmar Ratio            & 0.190 & 0.373$\pm$0.231 & -0.109$\pm$0.364 & 0.439$\pm$0.546 & 0.454$\pm$0.543 &  0.027$\pm$0.205 && \textbf{*0.828$\pm$0.543} & \\
Downside Deviation      & 0.109 & 0.125$\pm$0.039 &  0.173$\pm$0.098 & 0.132$\pm$0.081 & 0.115$\pm$0.029 &  0.122$\pm$0.022 && \textbf{*0.102$\pm$0.017} & \\
Max. Drawdown           & 0.276 & 0.248$\pm$0.084 &  0.443$\pm$0.153 & 0.285$\pm$0.115 & 0.272$\pm$0.097 &  0.356$\pm$0.113 && \textbf{*0.188$\pm$0.036} & \\
Avg. Profit / Avg. Loss & \textbf{*1.417} & 1.175$\pm$0.155 &  0.840$\pm$0.194 & 1.172$\pm$0.239 & 1.190$\pm$0.211 &  1.068$\pm$0.187 && 1.342$\pm$0.275 & \\
Hit Rate                & 0.435 & 0.499$\pm$0.030 &  0.481$\pm$0.035 & 0.505$\pm$0.026 & 0.500$\pm$0.029 &  0.472$\pm$0.019 && \textbf{*0.509$\pm$0.031} & \\
\bottomrule\end{tabular}
\end{table*}

Table \ref{tbl:trading_performance_comparison} consolidates trading performance across the various strategies after ten runs\footnote{The results of 1WR are obtained after a single run as there is no stochastic training involved. All other models are based on ten runs. 
We produce ten source models for each ranker that use transfer learning, i.e., SAR+ps and FEN. With each set, we pick the top 2 ranked by the Sharpe ratio and train five target models with each.
}.
While we compute the statistics here without transaction costs to focus on raw performance, we consider the impact of different cost assumptions later in \nameref{sec:turnover_analysis}. An additional layer of volatility scaling is applied at the portfolio level to facilitate comparison. Where applicable, each bold and asterisked figure indicate the best measurement across all models for its respective performance measures.

The statistics presented in Table \ref{tbl:trading_performance_comparison} demonstrate the superiority of the FEN model -- delivering an approximately threefold boosting of the Sharpe ratio relative to the heuristics-based 1WR and an improvement of about 50\% versus the best benchmark (SAR). From a risk perspective, FEN is among the least volatile and possesses the lowest downside deviation and max drawdown figures. We also note that the instability associated with cryptocurrencies is reflected across all models running at levels higher than the 15\% annual target.

Focusing on the reference baselines, SAR is the best model, while LN is the worst. While SAR's performance is likely a result of its sophisticated underlying model, it is only marginally better than the tree-based LM. This observation is reasonable since SAR and LM have produced state-of-the-art results on different LTR tasks \citep{pobrotynContextAwareLearningRank2020, qinAreNeuralRankers2021}. Additionally, the consequence of over-fitting is reflected in LN's results -- which are significantly worse than the simpler 1WR model. While both SAR and SAR+ps share the same transformer-based architecture, the latter's inferiority suggests that \textit{only} relying on a network pre-trained on FX data is insufficient in the context of this work.

In FEN, the pre-trained network runs in \textit{parallel} with the target encoder block. This setup leads to the model behaving like an ensemble of (potentially shallow) networks. Specifically, this setup avoids depending \textit{solely} on the pre-trained module -- which can be catastrophic for performance when affected by negative transfer (See \nameref{app:negative_transfer} in the Appendix) -- and instead allows extracted features to be flexibly combined in a data-driven manner. 

\subsection{Ranking Performance} \label{sec:ranking_performance}
We assess ranking precision with the NDCG (Normalised Discounted Cumulative Gain), a popular measure in the IR literature, with higher values implying greater accuracy in sorting objects (to match some optimal ordering). Given that we trade the top/bottom quintile each time, we concentrate on the average NDCG@2 for long and short trades aggregated over all weekly rebalances. Table \ref{tbl:ranking_performance_comparison} indicates that the MLP can rank the most accurately based on this measure, and this is followed by FEN. Although better accuracy generally implies higher returns and Sharpe ratios as seen in  Table \ref{tbl:trading_performance_comparison}, this does not hold for the MLP. 
This inconsistency can be resolved when we group FEN's returns conditioned on the relative accuracy (over rebalances) of FEN against the MLP\footnote{The MLP has a higher average NDCG@2 than FEN for about 52\% of the times when the portfolio is rebalanced.}. 
By segmenting returns and accuracy in this manner, we observe that FEN is only slightly worse off returns-wise when it is less accurate than the MLP but generates considerably higher returns when it is more accurate -- which explains its higher overall returns (and Sharpe ratio). These results are presented in Table \ref{tbl:segmented_returns_ndcg_comparison}.

\begin{table*}[t!]\centering
\caption{Ranking performance for all models based on the popular NDCG (Normalised Discounted Cumulative Gain). We calculate the average NDCG@2 over long and short positions across all rebalances. All figures except those belonging to 1WR are computed based on ten runs. Higher accuracy typically translates to higher returns and Sharpe ratios (See Table \ref{tbl:trading_performance_comparison}). FEN outperforms most benchmarks in terms of NDCG@2 except for the MLP.}
\label{tbl:ranking_performance_comparison}
\begin{tabular}{@{}rccccccccc@{}}
\toprule
& \multicolumn{6}{c}{\textbf{Benchmarks}}& \phantom{abc}& \multicolumn{2}{c}{\textbf{Proposed}} \\
\cmidrule{2-7}\cmidrule{9-10} & 1WR & MLP & LN & LM & SAR & SAR+ps && FEN \\
\midrule
NDCG@2  & 0.580$\pm$0.198 & \textbf{*0.615$\pm$0.007} & 0.597$\pm$0.010 & 0.601$\pm$0.007 & 0.604$\pm$0.010 & 0.594$\pm$0.007 && 0.607$\pm$0.011 \\
\bottomrule\end{tabular}
\end{table*}

\begin{table*}[t!]\centering
\caption{Segmented returns based on the relative ranking performances of the MLP and FEN models. We calculate the average NDCG@2 over long and short positions across all rebalances. The first two columns compare these measures at the times when FEN has an NDCG@2 lower (i.e., less accurate) than the MLP. The next pair of columns compare the same measures over instances when FEN has a higher NDCG@2. The magnitude of FEN's gains (when more accurate than the MLP) is larger than its losses (when less accurate).
}
\label{tbl:segmented_returns_ndcg_comparison}
\begin{tabular}{@{}rccccc@{}}
\toprule
& \multicolumn{2}{c}{\textbf{MLP more accurate}}& \phantom{abc}& \multicolumn{2}{c}{\textbf{FEN more accurate}}\\
\cmidrule{2-3}\cmidrule{5-6} & MLP & FEN && MLP & FEN \\
\midrule
NDCG@2          & 0.648$\pm$0.115 & 0.566$\pm$0.117 &&  0.576$\pm$0.108 & 0.648$\pm$0.105 \\
Avg. Returns    & 0.006$\pm$0.023 & 0.000$\pm$0.020 && -0.004$\pm$0.016 & 0.006$\pm$0.014 \\
\bottomrule\end{tabular}
\end{table*}

\subsection{Attention Heatmap Patterns}\label{sec:results_attention_weights}
\begin{figure*}[t!]
  \centering
  \includegraphics[scale=0.32]{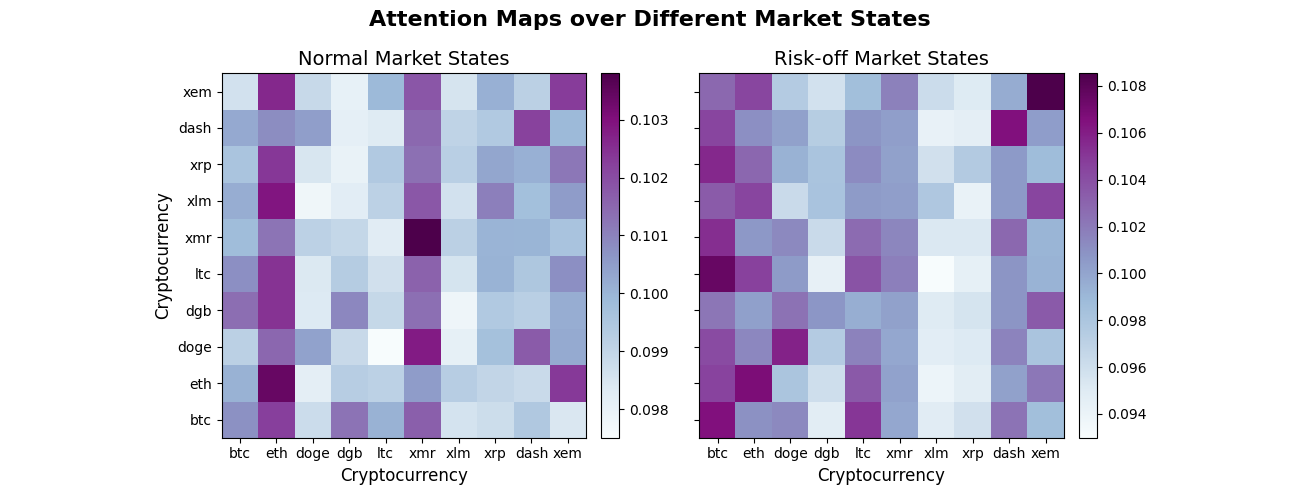}
  \caption{Aggregated attention heatmaps for the FEN model over both normal (left) and risk-off (right) periods. Localised groups of darker cells form across the map over normal states as measured by the VIX -- likely reflecting idiosyncratic developments within the set of cryptocurrencies. The behaviour of the attention map under risk-off regimes, during which multiple cryptocurrencies are affected, is reflected in the dispersion of darker-shaded cells. 
  }\label{fig:attn_map_broad_events}
\end{figure*}

Attention is an integral part of various state-of-the-art Transformer-based architectures, providing these models with an effective way to model relationships by attending to relevant regions in the data.
We examine the raw attention heatmaps of our proposed model to determine if this mechanism can learn helpful information.
Specifically, we visualise the \textit{last} attention layer of the model's target encoder stack since deeper layers tend to capture high-level representation and global relationships \citep{panLessMorePay2021}. 
Against increasingly coupled markets, volatility spillovers have grown increasingly frequent such as risk transmission from the equity to cryptocurrency markets \citep{uzonwanneVolatilityReturnSpillovers2021}.
Figure \ref{fig:attn_map_broad_events} depicts the aggregated heatmaps across normal and global risk-off regimes\footnote{We define a risk-off event at the weekly frequency as an instance when the daily VIX over any one or more days within the week is 5\% points higher than its 60-day moving average, and all normal states are thus defined to be non-risk-off states. This definition classifies approximately 20\% of the target dataset's evaluation period as risk-off.}. Darker cells correspond to larger weights which play a more significant role in the model's prediction\footnote{We also note that attention has been documented by \citet{serranoAttentionInterpretable2019} to be a noisy indicator of a model inputs' overall contribution to final prediction, and is not a fail-safe measure. Additionally, the heatmaps are not symmetric due to the softmax operator used in computing attention.}. 
The heatmaps show that the darker cells are more dispersed in risk-off states since multiple instruments are affected. This pattern of dispersion is less evident over normal states and is likely due to idiosyncratic developments in the cryptocurrency markets.  

\begin{figure*}[t!]
  \centering
  \includegraphics[scale=0.32]{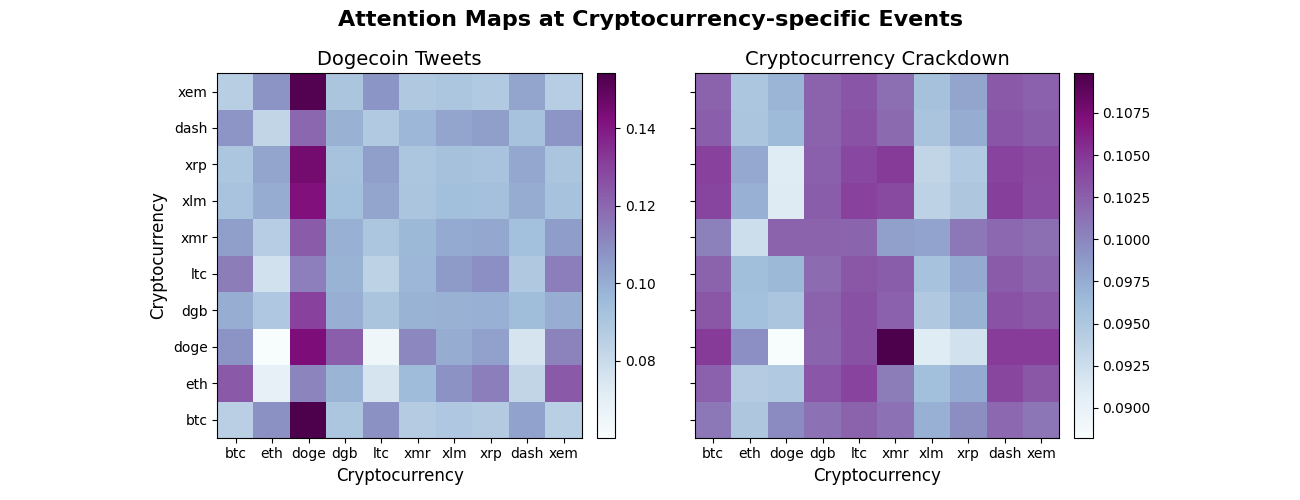}
  \caption{Aggregated attention heatmaps for the FEN model around the time of major cryptocurrency-related events. An activated column can be seen for Dogecoin around Elon Musk's tweets about the digital token during early May 2021 (left). The pattern of diffused shaded blocks in the second half of May 2021 coincides with China's government regulators declaring a crackdown on financial institutions' use of cryptocurrencies and reflects the risk-off sentiments inherent in this event (right).
  }\label{fig:attn_map_narrow_events}
\end{figure*}

Beyond the impactful macro-level episodes that cause sharp changes to the VIX, we observe similar behaviour of the FEN's attention mechanism at significant cryptocurrency-specific market events. In the left of Figure \ref{fig:attn_map_narrow_events} for instance, a distinct darker column of cells over Dogecoin formed around Elon Musk's tweet about the digital token \citep{westbrookMuskTweetsDogecoin2021}. On the right heatmap, the dispersed region of darker cells reflects the risk-off sentiments inherent in China's regulators' intent to crack down on the use of cryptocurrencies \citep{haleBitcoinGyratesFears2021}. These plots indicate that FEN's attention module is responsive to market dynamics and is capable of learning valuable information. More importantly, it appears to possess an interpretable structure and provide an additional perspective that helps explain the model's strong performance against other attention-based benchmarks. We refer the interested reader to the section titled \nameref{app:raw_attention_maps} in the Appendix to see how the heatmaps of other models compare.

\begin{table*}[t!]\centering
\caption{Average weekly portfolio turnover across all strategies. FEN has the lowest turnover.}
\label{tbl:turnover_comparison}
\begin{tabular}{@{}rccccccccc@{}}
\toprule
& \multicolumn{6}{c}{\textbf{Benchmarks}}& \phantom{abc}& \multicolumn{2}{c}{\textbf{Proposed}} \\
\cmidrule{2-7}\cmidrule{9-10} & 1WR & MLP & LN & LM & SAR & SAR+ps && FEN \\
\midrule
Turnover  & 0.832 & 0.710$\pm$0.030 & 0.711$\pm$0.047 & 0.787$\pm$0.018 & 0.676$\pm$0.038 & 0.715$\pm$0.053 && \textbf{*0.673$\pm$0.039} & \\
\bottomrule\end{tabular}
\end{table*}

\begin{table*}[t!]\centering
\caption{Sharpe ratios across selected profitable strategies with a 15\% annual volatility target under different transaction cost assumptions. Positive Sharpe ratios for FEN persists up to around 29 bps. This is above the 26 bps per trade assumed for cryptocurrencies and thus demonstrate their suitability for trading these instruments.}
\label{tbl:ex_cost_sharpe_comparison}
\begin{tabular}{@{}lccccccc@{}}
\toprule
& \multicolumn{4}{c}{\textbf{Benchmarks}}& \phantom{abc}& \multicolumn{2}{c}{\textbf{Proposed}} \\
\cmidrule{2-5}\cmidrule{7-8} Transaction Costs & 1WR & MLP & LM & SAR && FEN \\
\midrule
0 bps   & 0.238  &  0.403$\pm$0.240 & 0.456$\pm$0.546 & 0.496$\pm$0.583 && \textbf{*0.749$\pm$0.410} & \\
5 bps   & 0.115  &  0.273$\pm$0.248 & 0.302$\pm$0.542 & 0.360$\pm$0.580 && \textbf{*0.621$\pm$0.415} & \\
10 bps  & -0.011 &  0.144$\pm$0.257 & 0.153$\pm$0.532 & 0.225$\pm$0.576 && \textbf{*0.493$\pm$0.418} & \\
15 bps  & -0.140 &  0.016$\pm$0.267 &  0.11$\pm$0.515 & 0.089$\pm$0.574 && \textbf{*0.366$\pm$0.422} & \\
20 bps  & -0.271 & -0.111$\pm$0.276 & -0.134$\pm$0.506 & -0.046$\pm$0.571 && \textbf{*0.238$\pm$0.425} & \\
25 bps  & -0.404 & -0.236$\pm$0.286 & -0.283$\pm$0.504 & -0.180$\pm$0.569 && \textbf{*0.111$\pm$0.429} & \\
30 bps  & -0.538 & -0.359$\pm$0.297 & -0.433$\pm$0.504 & -0.315$\pm$0.567 && \textbf{*-0.017$\pm$0.432} & \\
\bottomrule\end{tabular}
\end{table*}

\begin{figure*}[t!]
  \centering
  \includegraphics[scale=0.5]{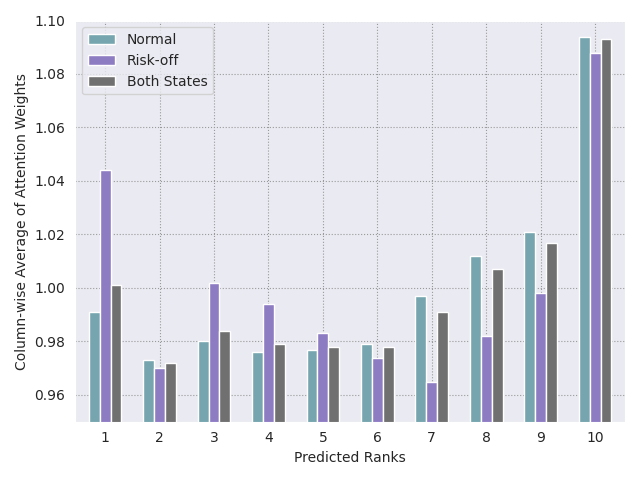}
  \caption{
    Column averages of the raw attention heatmaps for FEN across different market states aggregated over multiple runs. Each column average corresponds to a ranking index predicted by the model. Across the three different settings, the bars appear to roughly follow a 'J' shape, with higher weights for each half of the ranked list typically coinciding with traded positions (longs are labelled 9 and 10 on the horizontal axis, while shorts are 1 and 2). Values for the ranking index 2 diverge from this pattern.
  }\label{fig:fen_attn_hist}
\end{figure*}

To further examine the link between attention values, predicted ranks, and eventual trading positions, we compute the column-wise averages\footnote{
We compute column-wise averages as the softmax operation in Equation (\ref{eqn:attention}) is performed on each row of the matrix. Given this, if the model is paying more attention to a particular cryptocurrency, then the average of the heatmap cells in its column should register a higher value than the other column averages (For instance, refer to the discussion on the heatmap patterns formed around Elon Musk's tweet about Dogecoin and also the left heatmap in Figure \ref{fig:attn_map_narrow_events}).} of the attention heatmaps grouped by different market states and aggregated across all runs, with the final result presented in Figure \ref{fig:fen_attn_hist}. The histogram for each state has values that approximately follow a 'J' shape. Higher averages for the upper and lower halves of the ranked list generally coincide with trading positions (long trades are labelled 9 and 10 on the horizontal axis, while shorts are 1 and 2), with the longs' half of the histogram typically registering higher values. 
This pattern of higher weights near the ends of the ranked list is similarly observed in the other attention-based models SAR (Figure \ref{fig:tfr_attn_hist}) and SAR+ps (Figure \ref{fig:ps_attn_hist}) -- highlighting the close connection between attention, ranking outcomes, and, eventually, the traded positions generated by these models.

Before proceeding to the next section on turnover analysis, we highlight a limitation of the attention mechanism. Specifically, while it can pick up complex dependencies between the model's input elements, it is computationally prohibitive -- scaling quadratically with the length of the input sequence \citep{choromanskiRethinkingAttentionPerformers2021}. Hence, directly using FEN on a broader cross-sectional slate of instruments will likely be challenging. To this end, a few workarounds have been proposed \citep{belloAttentionAugmentedConvolutional2019, chanImputerSequenceModelling2020, childGeneratingLongSequences2019, gulatiConformerConvolutionaugmentedTransformer2020, beltagyLongformerLongDocumentTransformer2020}. We also note the recent work of \citet{choromanskiRethinkingAttentionPerformers2021} who develop Performers, which is not just linear in both time and space complexities but can produce provably accurate estimates of softmax full-rank attention.

\subsection{Turnover Analysis} \label{sec:turnover_analysis}
To investigate the impact of transaction costs on strategy performance, we first compute the weekly portfolio turnover at period $t$ which we define as:
\begin{equation}
    \zeta^{(i)}_t = \sigma_{tgt} \sum_{i=1}^n \Biggl | \frac{S^{(i)}_t}{\sigma^{(i)}_t} - \frac{S^{(i)}_{t-1}}{\sigma^{(i)}_{t-1}} \Biggr |
\end{equation}
Table \ref{tbl:turnover_comparison} shows the FEN model as among the strategies with the lowest average turnover over 2018-2021. To further examine and quantify the impact of transaction cost on performance, we also compute the ex-cost Sharpe ratios and present the results in Table \ref{tbl:ex_cost_sharpe_comparison}. The figures show that FEN can deliver higher and positive cost-adjusted Sharpe ratios up to 29 bps -- above the 26 bps per trade assumed for cryptocurrencies \citep{tzouvanasMomentumTradingCryptocurrencies2020} and therefore demonstrating its suitability for trading these instruments.

\section{Conclusion}
We introduce Fused Encoder Networks (FEN) -- a novel hybrid parameter-sharing transfer ranking model that overcomes the difficulties of constructing cross-sectional strathttps://www.ft.com/content/29d7826c-b591-4c02-aa13-d570f5629c11egies based on neural rankers to trade instruments with short histories. Using the popular momentum strategy applied to cryptocurrencies as a demonstrative use-case, we validate the effectiveness of our proposed approach. Concretely, the FEN surpasses the reference baselines, which include state-of-the-art LTR algorithms, on most performance measures -- notably delivering an approximately threefold boosting of the Sharpe ratio over risk-adjusted momentum and a gain of approximately 50\% over the best benchmark. Future work includes modifying various components of the original framework, such as introducing additional processing in the output head or replacing the listwise loss with other loss functions. Another direction is to explore transferring and synthesising knowledge from other/multiple source domains.

\begin{acks}
We would like to thank the Oxford-Man Institute of Quantitative Finance for financial and computing support. SR is grateful for support from the Royal Academy of Engineering.
\end{acks}

%
\bibliographystyle{ACM-Reference-Format}
\bibliography{refs}


\appendix

\section{Learning to Rank and Transfer Ranking}\label{app:transfer_ranking_framework}

\subsection{Learning to Rank for Cross-Sectional Momentum} \label{app:ltr_setup}

Learning to Rank (LTR) is a supervised learning problem where the goal is to sort a list of objects to maximise some measure of utility (e.g., discounted cumulative gain). For training, we are presented with a set of dates $\mathcal{B} = \{b_1, ... b_{t-1}\}$ where at each $b_i$ we have $n$ instruments $\textbf{c}_i = \{c_i^{(1)}, ... c_i^{(n)}\}$. We also have a corresponding set of quintiles $\mathcal{Q}$ of realised returns $\textbf{q}_{i+1}$ from the \textit{next} period that $\textbf{c}_i$ have been assigned based on an ascending sort. Here,  $\textbf{q}_{i+1}=\{q_{i+1}^{(1)}, ... , q_{i+1}^{(n)}\}$ where $q_{i+1}^{(j)} \in \mathcal{Q} = \{Q_1, ... , Q_5\} \textrm{ for } \forall \textrm{ } i, j$. Additionally, $k > l \Rightarrow Q_k \succ Q_j \textrm{ for } k, l \in \{1, ..., 5\}$ where $\succ$ stands for the order relation. In other words, instruments possessing higher realised returns are assigned higher quintiles. With each rebalance-instrument pair at period $i$, a feature vector $u^{(j)}_i = \phi(b_i, c^{(j)}_i)$ can be formed, noting that $\phi(\cdot)$ characterises the feature function and that $i \in \{1, ..., t\} \textrm{ and } j \in \{1, ..., n\}$. Letting $\textbf{u}_i = \{u^{(1)}_i, ..., u^{(n)}_i\}$, we can go on to form the broader training set $\{\textbf{u}_i, \mathbf{q}_{i+1} \}^{t-1}_{i=1}$. The goal of LTR is to learn a function $f$ that predicts a score $f(u^{(j)}_{t}) = z^{(j)}_{t}$ at the Score Calculation phase (refer to the Traditional LTR pipeline in Figure \ref{fig:pipeline}) when presented with an out-of-sample input $u^{(j)}_{t}$. For more details, we point the reader to \citet{pohBuildingCrossSectionalSystematic2021} and \citet{liLearningRankInformation2011}.

\subsection{From Learning to Rank to Transfer Ranking}

Transfer Ranking (TR) builds on top of the LTR framework. A \textit{domain} $\mathcal{D}$ consists of a feature space $\mathcal{X}$ and a marginal distribution $P(X)$, i.e., $\mathcal{D}=\{\mathcal{X}, P(X)\}$. $X$ represents an instance set and is defined as $X = \{\textbf{x}|x_i \in \mathcal{X}, i = 1, ..., n \}$. A \textit{task} $\mathcal{T}$ is composed of a label space $\mathcal{Y}$ and a decision function $f$, so $\mathcal{T}=\{\mathcal{Y}, f \}$. The decision function $f$ is learned from the data. 
Using these definitions, the goal of LTR for CSM in 
the preceding section
can be expressed as learning the ranking function $f$ in $\mathcal{T}=\{\mathcal{Q}, f \}$ over some domain of interest $\mathcal{D}=\{\mathcal{U}, P(U)\}$. The feature space and instance set are represented by $\mathcal{U}$ and $U$ respectively. A single instance $\textbf{u}_i$ at period $i$ is given as
$\textbf{u}_i = \{u^{(1)}_i, ..., u^{(n)}_i\}$ where $u^{(j)}_i \in \mathcal{U} \textrm{ for } \forall \textrm{ } i, j$.

In the TR setting, we are given observations corresponding to $m^S \in \mathbb{N}^+$ \textit{source} domains and tasks, i.e., $\bigl \{(\mathcal{D}_{S_i}, \mathcal{T}_{S_i}) \textrm{ | } i = 1, ..., m^S \bigr \}$ as well as observations about $m^T \in \mathbb{N}^+$ \textit{target} domains and tasks, i.e., $\bigl \{(\mathcal{D}_{T_j}, \mathcal{T}_{T_j}) \textrm{ | } j = 1, ..., m^T \bigr \}$. TR aims to exploit the knowledge embedded in the source domains to enhance the performance of the learned decision functions $f_{T_j}, j=1, ..., m^T$ on the target domains. 
In this work, we focus on $m^S =m^T = 1$. Specifically, we want to enhance the performance of $f_T$, which is the scoring function for cryptocurrencies, by incorporating knowledge extracted from the FX dataset $f_S$. This is achieved by sharing the parameters $\theta_S$ of the calibrated $f_S$ with $f_T$ as described in \nameref{sec:problem_definition}.

\section{Dataset Details}\label{app:data}
\subsection{Source Dataset}
Our source dataset is composed of 31 currencies all expressed versus the USD over the period 2-May-2000 to 31-Dec-2021.
We make use of this data to train the upstream model required for knowledge transfer. The full list of currencies is as follows:
\small
\begin{itemize}
    \item G10: AUD, CAD, CHF, EUR, GBP, JPY, NOK, NZD, SEK, USD
    \item EM Asia: HKD, INR, IDR, KRW, MYR, PHP, SGD, TWD, THB
    \item EM Latam: BRL, CLP, COP, PEN, MXN
    \item CEEMEA: CZK, HUF, ILS, PLN, RUB, TRY
    \item Africa: ZAR
\end{itemize}
\normalsize
\subsection{Target Dataset}
Our target dataset is focused on the following 10 cryptocurrencies: Bitcoin (BTC), Ethereum (ETH), DogeCoin (DOGE), DigiByte (DGB), Litecoin (LTC), Stellar (XLM), XRP,  Monero (XMR), NEM (XEM), Dash (DASH). The dataset for these digital tokens span 1-Jan-2016 to 31-Dec-2021.

\subsection{Data Preprocessing}
In order to reduce the impact of outliers during training, we winsorise both foreign exchange and cryptocurrency data by capping and flooring it to be within 3 times its exponentially weighted moving (EWM) standard deviation from its EWM average that is calculated with a 26-week halflife decay.

\section{Additional Training Details}\label{app:training_details}

\textit{Model Fine-tuning:} We fine-tune both FEN and SAR+ps models. In this step, we unfreeze the entire network and re-train with the lowest value in the learning rate search range, i.e., $10^{-6}$ and over different batch sizes.

\textit{Python Libraries:} LambdaMART uses {\tt XGBoost} \citep{chenXGBoostScalableTree2016}. All other machine learning models are developed with {\tt Tensorflow} \citep{abadiTensorFlowLargescaleMachine2015}.

\textit{Hyperparameter Optimisation:} Hyperparameters are tuned using {\tt HyperOpt} \citep{bergstraHyperoptPythonLibrary2015}. For LambdaMART, we refer to the hyperparameters as they are named in the {\tt XGBoost} library. The feature search spaces for various machine learning models are as follows:

\vspace{8pt}
\textbf{Multi-layer Perceptron (MLP):}
\begin{itemize}
    \item Dropout Rate -- [0.0, 0.2, 0.4, 0.6, 0.8]
    \item Hidden Width -- [8, 16, 32, 64, 128]
    \item Batch Size -- [8, 16, 32, 64, 128]
    \item Learning Rate -- [$10^{-6}, 10^{-5}, 10^{-4}, 10^{-3}, 10^{-2}, 10^{-1}]$
\end{itemize}

\vspace{8pt}
\textbf{ListNet (LN):}
\begin{itemize}
    \item Dropout Rate -- [0.0, 0.2, 0.4, 0.6, 0.8]
    \item Hidden Width -- [8, 16, 32, 64, 128]
    \item Batch Size -- [1, 2, 4, 8, 16]
    \item Learning Rate -- $[10^{-6}, 10^{-5}, 10^{-4}, 10^{-3}]$
\end{itemize}

\vspace{8pt}
\textbf{LambdaMART (LM):}
\begin{itemize}
    \item `objective' -- `rank:pairwise'
    \item `eval\_metric' -- `ndcg'
    \item `num\_boost\_round' -- [50, 100, 250, 500, 1000]
    \item `max\_depth' -- [6, 7, 8, 9, 10]
    \item `eta' -- $[10^{-6}, 10^{-5}, 10^{-4}, 10^{-3}, 10^{-2}, 10^{-1}]$
    \item `reg\_alpha' -- $[10^{-6}, 10^{-5}, 10^{-4}, 10^{-3}, 10^{-2}, 10^{-1}]$
    \item `reg\_lambda' -- $[10^{-6}, 10^{-5}, 10^{-4}, 10^{-3}, 10^{-2}, 10^{-1}]$
    \item `tree\_method' -- `gpu\_hist'
\end{itemize}

\vspace{8pt}
\textbf{Self-attention Ranker (SAR), Self-attention Ranker with Parameter-sharing (SAR+ps), both the source model for Fused Encoder Networks (FEN) and FEN itself:}
\begin{itemize}
    \item $d_{fc}$ -- [8, 16, 32, 64, 128]
    \item $d_{ff}$ -- [8, 16, 32, 64, 128]
    \item Dropout Rate -- [0.0, 0.2, 0.4, 0.6, 0.8]
    \item Hidden Width -- [16, 32, 64, 128, 256]
    \item Batch Size -- [2, 4, 6, 8, 10]
    \item Learning Rate -- $[10^{-6}, 10^{-5}, 10^{-4}, 10^{-3}]$
    \item No. of Encoder Layers -- [1, 2, 3, 4]
    \item No. of Attention Heads -- 1
\end{itemize}

\vspace{8pt}
\textbf{Fine-tuning step for SAR+ps and FEN:}
\begin{itemize}
    \item Batch Size -- [2, 4, 6, 8, 10]
    \item Learning Rate -- $10^{-6}$
\end{itemize}

\section{Incorporating Knowledge from Sub-optimal Source Models}\label{app:negative_transfer}
We investigate the impact on model performance when inappropriate source models are used for knowledge transfer. Out of 10 runs of the source model, we pick the \textit{worst} 2 as measured by the strategy's Sharpe ratio on the out-of-sample period. Using components from each, we construct their respective target models and make predictions on the corresponding target dataset. The results of this exercise are presented in Table \ref{tbl:negative_transfer_performance}. Importantly, we see that transferring knowledge from sub-optimal source models has a substantial performance drag for FEN (around 45\% reduction in the Sharpe ratio). 

\begin{table}[t!]\centering
\caption{Performance metrics for FEN and SAR+ps over the 2018-2021 period with a 15\% annual volatility target. All figures are based on ten runs. Both models incorporate components from \textit{sub-optimal} source models. The impact is clearly evident for both models with statistics that are worse off than their respective counterparts in Table \ref{tbl:trading_performance_comparison}.}
\label{tbl:negative_transfer_performance}
\begin{tabular}{@{}rccc@{}}
\toprule
 & SAR+ps & FEN & \\
\midrule
E[returns] 	                & -0.050$\pm$0.097 & 0.078$\pm$0.055 & \\
Volatility 	                &  0.174$\pm$0.013 & 0.190$\pm$0.018 & \\
Sharpe Ratio 	            & -0.277$\pm$0.569 & 0.414$\pm$0.302 & \\
Sortino Ratio 	            & -0.260$\pm$0.819 & 0.719$\pm$0.527 & \\
Calmar Ratio 	            & -0.009$\pm$0.321 & 0.364$\pm$0.264 & \\
Downside Deviation 	        &  0.131$\pm$0.025 & 0.117$\pm$0.018 & \\
Max. Drawdown 	            & -0.424$\pm$0.173 & 0.240$\pm$0.066 & \\
Avg. Profit / Avg. Loss     &  1.016$\pm$0.249 & 1.189$\pm$0.163 & \\
Hit Rate 	                &  0.459$\pm$0.026 & 0.496$\pm$0.020 & \\
\bottomrule\end{tabular}
\end{table}

\section{Comparing Attention Heatmaps}\label{app:raw_attention_maps}
This section compares the heatmaps of models that use the attention mechanism: the SAR, SAR+ps, and FEN. Specifically, we present the weights of each model's last attention layer for each market state and cryptocurrency-specific event discussed in \nameref{sec:results_attention_weights}. Darker cells correspond to larger weights, contributing more to the model's prediction. Additionally, attention weights across different models are \textit{not} comparable since these are raw weights. Overall, the patterns produced by FEN appear to broadly align with what one might expect. Risk-off states, for instance, are associated with heatmaps with darker cells that are dispersed -- reflecting the broad impact felt by multiple instruments. This development is observed to a lesser extent in SAR, and is almost entirely absent in SAR+ps.

\begin{figure*}
  \centering
  \includegraphics[scale=0.30]{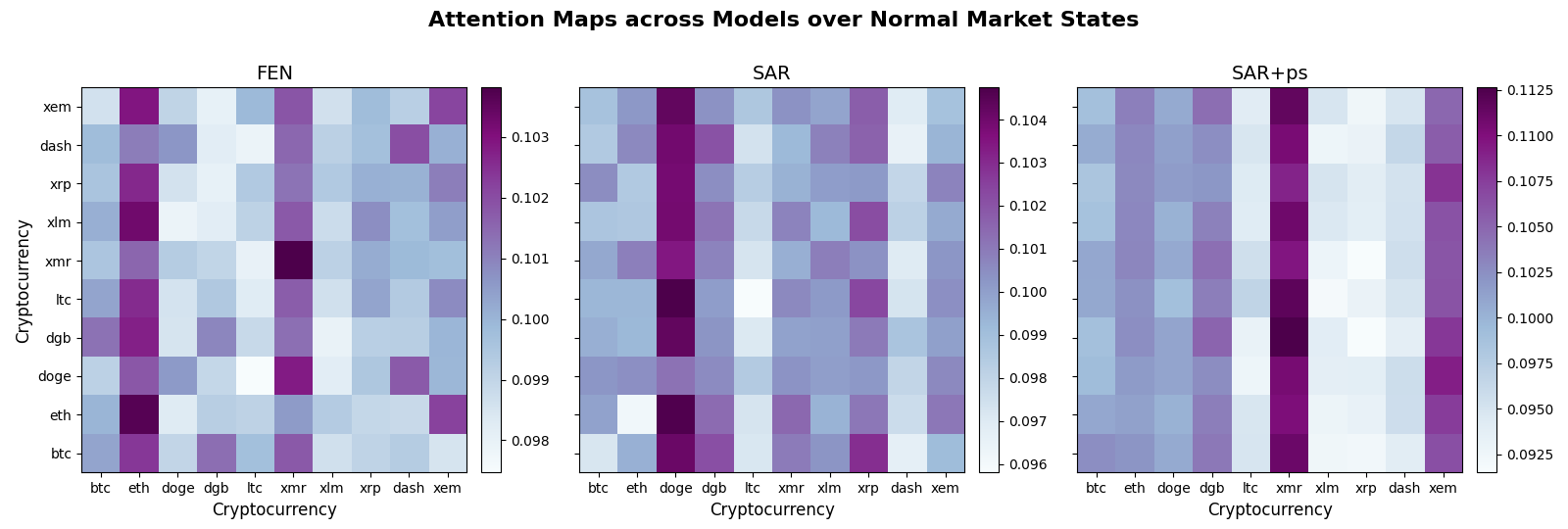}
  \caption{Aggregated maps over normal market states. Cells with larger weights cluster in pockets for FEN, but are dispersed for SAR. SAR+ps interestingly places a high amount of attention weights on Monero ('xmr') and NEM ('xem'), which are relatively less popular than Bitcoin ('btc') and Ethereum ('eth').
  }\label{fig:attn_map_comparison_normal}
\end{figure*}

\begin{figure*}
  \centering
  \includegraphics[scale=0.30]{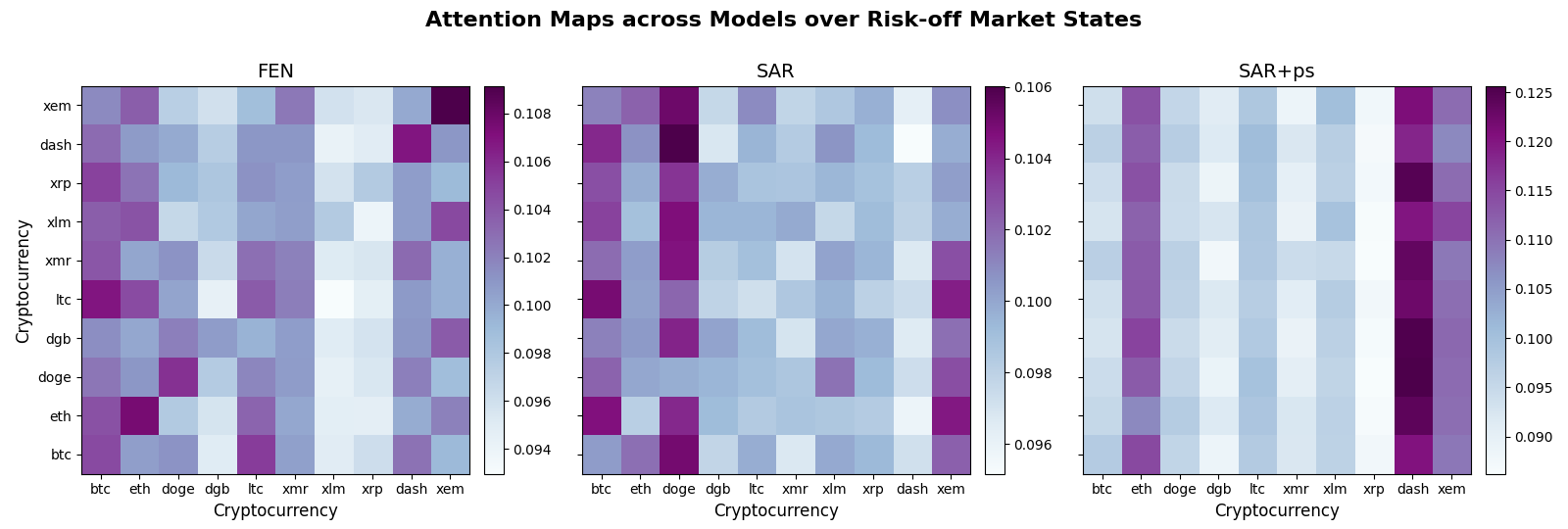}
  \caption{Aggregated maps over risk-off market states. Darker cells are dispersed more in FEN and SAR, but less for SAR+ps.
  }\label{fig:attn_map_comparison_riskoff}
\end{figure*}
\begin{figure*}
  \centering
  \includegraphics[scale=0.30]{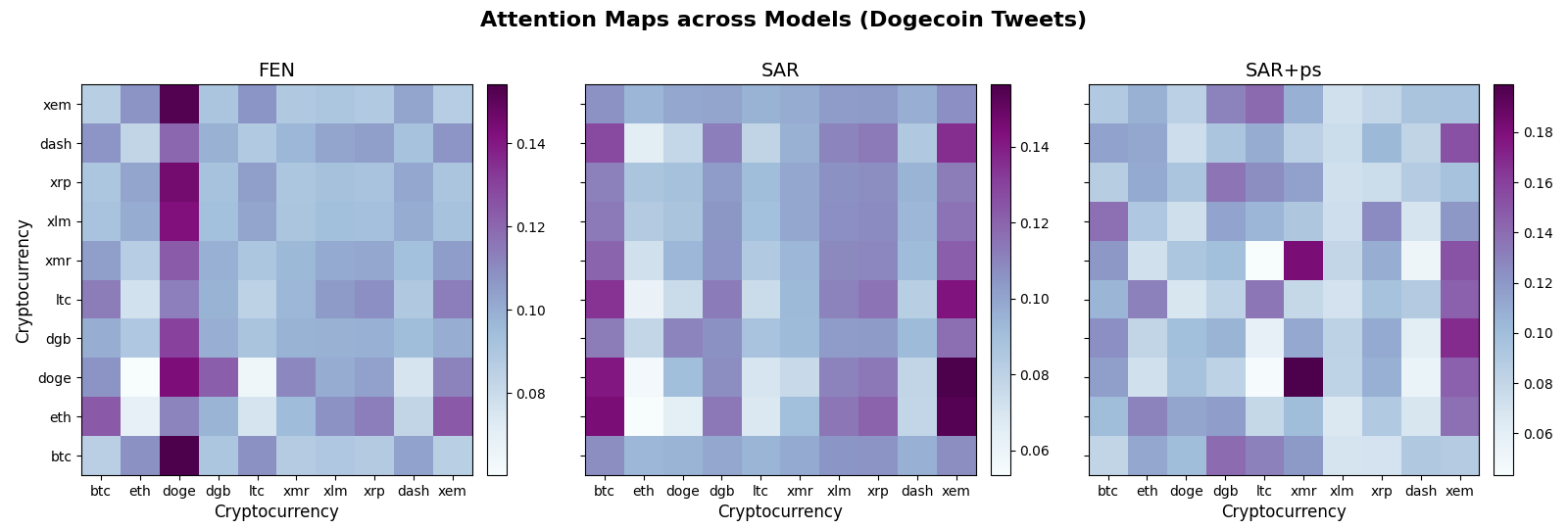}
  \caption{Attention maps in early May 2021, around the time of Elon Musk's Dogecoin ('doge') tweets. The 'doge' column is activated for FEN, but this pattern is absent for both SAR and SAR+ps.
  }\label{fig:attn_map_comparison_doge}
\end{figure*}

\begin{figure*}
  \centering
  \includegraphics[scale=0.30]{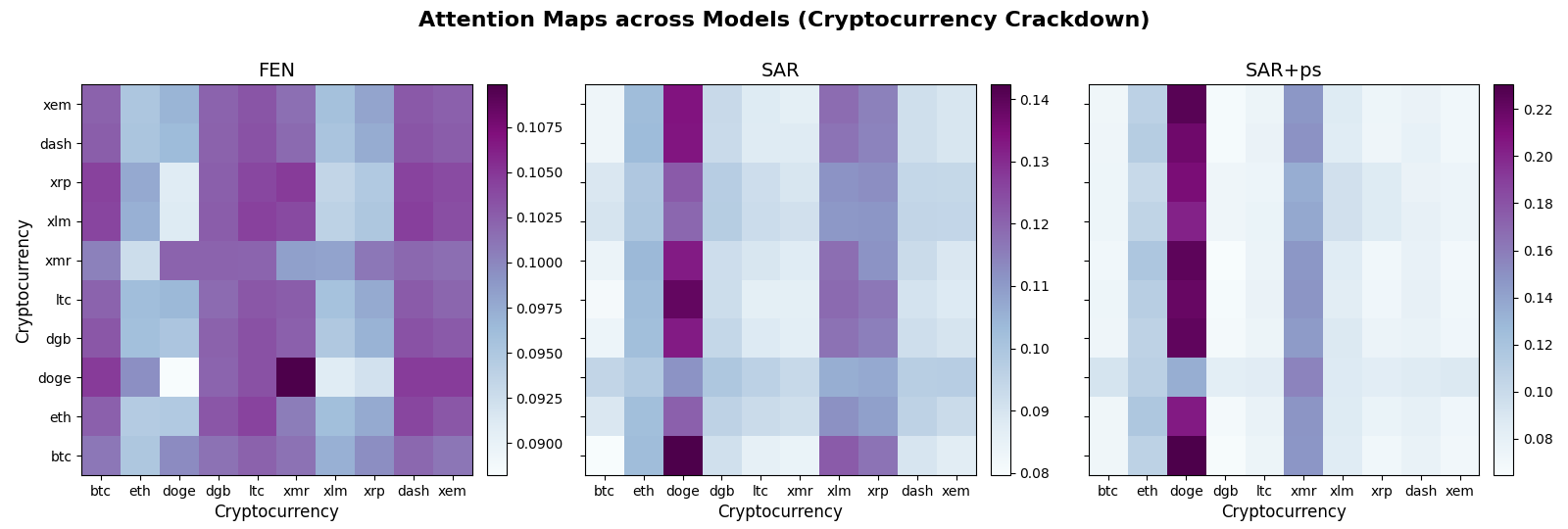}
  \caption{Attention maps for the second half of May 2021,  which covers China's government regulators declaring a crackdown on the use of cryptocurrencies by domestic financial institutions on the 19th of the same month. The significance of this event is evident from the pattern of diffused shaded blocks for FEN, with multiple coins affected. This is observed to a lesser extent in SAR's attention map and is almost completely absent in SAR+ps's.
  }\label{fig:attn_map_comparison_crypto}
\end{figure*}

\begin{figure*}
  \centering
  \includegraphics[scale=0.50]{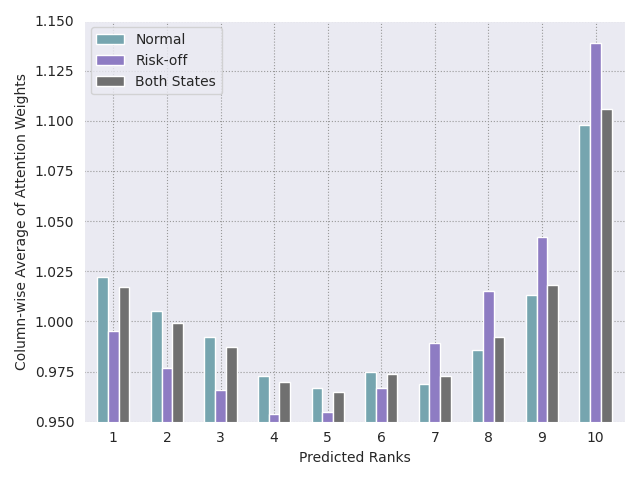}
  \caption{
  Column averages of the raw attention heatmaps for SAR across different market states aggregated over multiple runs. Each column average corresponds to a ranking index predicted by the model. Across the three different settings, the bars appear to roughly follow a 'J' shape, with higher weights for each half of the ranked list typically coinciding with traded positions (longs are labelled 9 and 10 on the horizontal axis, while shorts are 1 and 2). 
  }
  \label{fig:tfr_attn_hist}
\end{figure*}

\begin{figure*}
  \centering
  \includegraphics[scale=0.50]{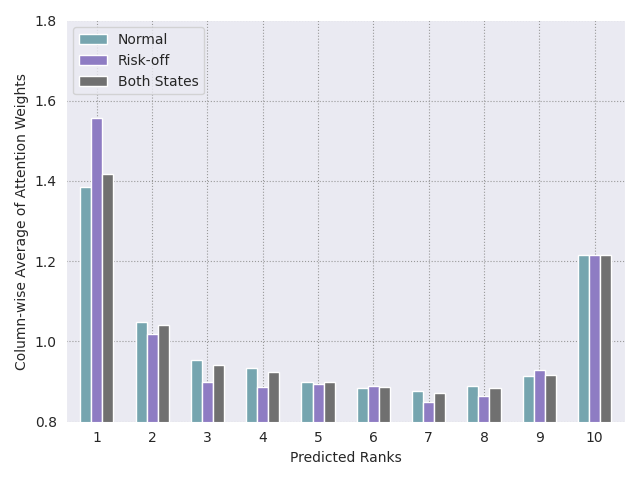}
  \caption{
  Column averages of the raw attention heatmaps for SAR+ps across different market states aggregated over multiple runs. Each column average corresponds to a ranking index predicted by the model. Across the three different settings, the bars appear to roughly follow a reversed 'J' shape flipped about the y-axis, with higher weights for each half of the ranked list typically coinciding with traded positions (longs are labelled 9 and 10 on the horizontal axis, while shorts are 1 and 2). 
  }
  \label{fig:ps_attn_hist}
\end{figure*}

\end{document}